\documentclass{aa}  
\usepackage{amsmath}
\usepackage{graphicx}
\usepackage{appendix}
\usepackage{multirow}
\usepackage[varg]{txfonts}
\usepackage{natbib}
\usepackage{array}
\usepackage{nicefrac}
\bibpunct{(}{)}{;}{a}{}{,}
\usepackage[version=4]{mhchem}
\usepackage{soul}
\usepackage{threeparttable}
\usepackage[dvipsnames]{xcolor}
\usepackage{ulem}
\usepackage{comment}
\usepackage[breaklinks,pdfnewwindow=true,colorlinks=true,citecolor=blue]{hyperref}

\inputencoding{utf8}

\definecolor{orange}{rgb}{1,0.5,0}

\definecolor{pink}{rgb}{0.858, 0.188, 0.478}

\definecolor{darkgreen}{rgb}{0,0.5,0}

\definecolor{lightgreen}{rgb}{0,0.5,0}

\begin{document}

\title{Broadband spectroscopy of astrophysical ice analogues}
\subtitle{IV. Optical constants of N$_2$ ice in the terahertz and mid-infrared ranges}

\author{
    F.~Kruczkiewicz \inst{\ref{MPE}, \ref{AMU}}\thanks{Present address: Leiden Observatory, Leiden University, Niels Bohrweg 2, 2333 CA Leiden, The Netherlands},
    A.A.~Gavdush \inst{\ref{GPI}},
    F.~Ribeiro \inst{\ref{MPE}, \ref{IFRJ}},
    D.~Campisi\inst{\ref{ITC}, \ref{DE}}\thanks{Present address: Aarhus Institute of Advanced Studies (AIAS), Aarhus University, Høegh-Guldbergs Gade 6B, 8000 Aarhus C, Denmark; Center for Interstellar Catalysis, Department of Physics and Astronomy, Aarhus University, 1520-337 Ny Munkegade 120, 8000 Aarhus C, Denmark},
    A.~Vyjidak \inst{\ref{MPE}},
    B.M.~Giuliano\inst{\ref{MPE}},
    G.A.~Komandin \inst{\ref{GPI}},
    S.V.~Garnov \inst{\ref{GPI}},
    T.~Grassi \inst{\ref{MPE}},
    P.~Theulé \inst{\ref{AMU}},
    K.I.~Zaytsev \inst{\ref{GPI}},
    A.V.~Ivlev \inst{\ref{MPE}},
    and P.~Caselli \inst{\ref{MPE}}
}

\institute{Max-Planck-Institut f\"ur Extraterrestrische Physik,
Gießenbachstraße 1, Garching, 85748, Germany
\label{MPE}
\and 
Aix Marseille Univ, CNRS, CNES, LAM, Marseille, France 
\label{AMU}
\and 
Prokhorov General Physics Institute of
the Russian Academy of Sciences, 
119991 Moscow, Russia
\label{GPI}
\and
Institute for Theoretical Chemistry, University of Stuttgart, Pfaffenwaldring 55,  Stuttgart, Germany
\label{ITC}
\and
Department of Engineering, University of Perugia, Via Duranti 93, 06125 Perugia, Italy
\label{DE}
\and
Federal Institute of Education, Science and Technology of Rio de Janeiro (IFRJ), Nilópolis Campus - Rio de Janeiro - Brazil
\label{IFRJ}
}

\date{Received - , 2024;
Accepted - , 2024}
   
\authorrunning{F.~Kruczkiewicz~et~al.}

\titlerunning{Dielectric spectroscopy of
\ce{N2} ices in the THz--IR range}

\date{Received 2024; accepted  2024}

    
\abstract
   {
   Understanding the optical properties of
   astrophysical ices is crucial for
   modeling dust continuum emission
   and radiative transfer in
   dense, cold interstellar environments.
   Molecular nitrogen,
   a primary carrier of N in protoplanetary disks,
   plays a key role in the formation of
   nitrogen-bearing species.
   However, the lack of direct measurements of
   the terahertz (THz)--infrared (IR)
   optical constants of \ce{N2} ice
   introduces uncertainties in
   radiative transfer models,
   snowline locations,
   and disk mass estimates.}
   {We present direct measurements and analysis of
   the optical properties of \ce{N2} ice
   across a broad THz--IR spectral range by
   combining THz pulsed spectroscopy (TPS) and Fourier-transform IR
   spectroscopy (FTIR).
   The observed optically active
   THz vibrational modes of \ce{N2} ice
   are supported by
   density functional theory (DFT) calculations.
   The consistency of our measurements and calculations
   with data sets from the literature is also assessed.}
   {\ce{N2} ice was grown at cryogenic temperatures
   via the gas-phase deposition onto a cold silicon window.
   The optical properties of the ice samples
   were quantified using our earlier-reported method:
   it involves the direct reconstruction of the
   THz complex refractive index from the
   TPS data, combined with the derivation of IR response from the FTIR data using the Kramers-Kronig relations. The \ce{N2} ice response was parameterized using the Lorentz
   model of complex dielectric permittivity, which was verified with our DFT calculations and compared with the literature data.}
    {The complex refractive index of \ce{N2} ice is quantified
    in the frequency range of $\nu = 0.3$--$16$~THz
    (the wavelength range of $\lambda = 1$~mm--$18.75$~$\mu$m),
    and compared with the DFT results
    as well as with the available literature data.
    The observed resonant absorption peaks
    at $\nu_\mathrm{L} = 1.47$ and $2.13$~THz,
    with the damping constants of
    $\gamma_\mathrm{L} = 0.03$ and $0.22$~THz,
    respectively, are
    attributed to the well-known optically active phonons
    of the $\alpha$-\ce{N2} crystal.}
    {We provide a complete set of the THz--IR optical constants for \ce{N2} ice by combining TPS and FTIR spectroscopy. Our results have implications for future observational and modeling studies of protoplanetary disk evolution and planet formation.}
    

\maketitle

\section{Introduction}

The distribution and abundance of ice components in
protoplanetary disks are crucial for
shaping the properties of emerging planetary systems.
Ices influence both
the growth and the inward migration of dust grains
-- the fundamental building blocks of planets,
while also dictating the reservoir of
key volatile elements
\citep{Gundlach2015Stickiness, Drazkowska2017Planetesimal, Sturm2023}.
The ice composition plays an important role in the thermodynamics and mass transfer in a disk, thus affecting the snowlines' location and the efficiency of planetesimal formation
\citep{Oberg2011Effectsofsnowlines, Arabhavi2022, Gavino2023}.
Measurements and analysis of
dust and ice opacities
are powerful tools for better constraining
the physical and chemical properties of disks.
While the underlying optical constants are known,
observations of dust continuum emission
and self-consistent modeling of dust opacity
provide refined estimates of
the disk mass, composition, and snowlines,
as compared to those coming
from models based on the bare-grain assumption
\citep{Arabhavi2022}.
The key to realizing these improvements is obtaining
reliable optical constants of ices
across a wide terahertz (THz)--infrared (IR) range
relevant to protoplanetary disks.

Traditionally, ice optical constants
have been derived using
the Fourier-transform IR (FTIR) spectroscopy,
which yields only
the amplitude transmission spectra
and which is commonly aided by
the Kramers-Kronig relations
to retrieve the phase for
the complete characterization of an ice sample
\citep{Bergren1978OHstretching, Hagen1981InfraredSpectrum, Hudgins1993}.
Although this approach
has led to invaluable catalogs of optical data, reliance on
the underlying assumptions and mathematical derivations
can introduce measurement uncertainties.
To mitigate this difficulty,
the THz pulsed spectroscopy (TPS) technique was applied by
\cite{AA.629.A112.2019}
to measure the THz response of ices
with no need for the Kramers-Kronig transform.
Indeed, TPS offers a favorable opportunity to
detect the THz waveform
transmitted through an analyte,
the Fourier transform, 
which gives spectral amplitude and phase.
This makes it possible to directly retrieve
the THz complex refractive index 
in a broad spectral range.
It was shown by
\cite{2022A&A...667A..49G}
that the spectral phase of TPS
can be used to calibrate that of FTIR data,
and thus to eliminate
the Kramers-Kronig transform uncertainty, facilitating
the broadband characterization of laboratory ices.
Such a technique was then applied
to comprehensively analyze
the THz--IR optical constants of
\ce{CO} and \ce{CO2} ices
\citep{2022A&A...667A..49G},
as well as their scattering properties and porosity
\citep{AA.scattering.2025}
relying on the broadband optical spectra.

The next goal of our broadband spectroscopic measurements is the
characterization of \ce{N2} ice.
Among all volatiles, molecular nitrogen is recognized
as a primary nitrogen carrier in the protoplanetary disks.
It plays a crucial role in the formation of abundant N-bearing gas phase species such as \ce{NH3} and \ce{N2H+} \citep[e.g.][]{AA.513A.41H.2010}.
In our Solar System, the presence of \ce{N2} ice has been confirmed in dwarf planets Pluto and Eris and Neptune’s largest moon Triton \citep{ApJ.725.1296T.2010, ApJ.751.76T.2012}. 

Because \ce{N2} has no strong dipole moment,
it is notoriously difficult to observe directly,
as it lacks strong optically active
vibrational modes. Estimates of its abundance depend on
the indirect tracers,
such as \ce{N2H+}, the CN/HCN ratio, and various nitriles
\citep{Qi_2019, van_t_Hoff_2017}.
These proxies provide
valuable constraints on the \ce{N2} distribution,
especially in cold outer disks,
where \ce{N2} is expected to freeze out
once temperatures drop below $20$--$25$~K
\citep{Minissale2022}.
However, indirect methods inevitably introduce
observational and chemical modeling uncertainties,
highlighting the need for complementary approaches,
particularly in the solid phase.
Understanding the ice reservoir of \ce{N2}
is important not only to constrain
its overall abundance but also to elucidate
its role in the formation of species
bearing \ce{N} on grains in the densest disk regions
\citep{Walsh2014}.

In this study, \ce{N2} ice is grown at cryogenic temperatures
and measured using TPS and FTIR techniques.
The optical properties of \ce{N2} ice
are retrieved in a broad spectral range
spanning the frequencies of $\nu = 0.3$--$16$~THz,
or the wavelengths of
$\lambda \simeq 1$~mm--$18.75$~$\mu$m.
The observed optically active vibrational modes 
are attributed to the quadrupole moments of
a molecule in a crystalline lattice,
parametrized by the Lorentz model,
and supported by density functional theory (DFT) calculations
and available literature data.
The obtained data on
the THz--IR optical properties of \ce{N2} ice
would be useful for modeling nitrogen freeze-out
and chemistry in protoplanetary disks.
Our findings can be used to calculate how the opacities of dust grains change when they are covered in ice mantles, particularly in spectroscopic regions where the ices' resonance features are present. This information aids in the interpretation of dust continuum emission, including variations expected between dense and diffuse regions where the fraction of ice-coated grains differs.

\section{Methods}
\label{SEC:ExpMeth}

\subsection{The experimental setup}
\label{SEC:Cryo}

The experimental studies
were conducted at the CASICE laboratory
developed at the Center for Astrochemical Studies
at the Max Planck Institute for Extraterrestrial Physics
(Garching, Germany).
Data obtained with the TPS
\citep{AA.629.A112.2019}
and FTIR
\citep{Mueller+2018}
spectrometers were combined to derive
the broadband optical constants of \ce{N2} ice.

In the original design of our laboratory,
the same vacuum chamber can be coupled
to both the TPS and the FTIR spectrometers.
A motorised translation stage moves the chamber
between the sample compartments of
the two spectrometers
and adjusts the position relative to the beam.
This ensures that the ice samples
analyzed with the two instruments
have reproducible properties
for given deposition conditions.
Therefore, the spectroscopic data
recorded in the THz and IR ranges
can be directly merged for the calculation of
the broadband complex refractive index of an ice. However, the measurements have not been performed on an identical set of ice samples.

A scheme of the vacuum chamber,
TPS and FTIR arrangements are detailed in
\cite{2022A&A...667A..49G}.
The $15$~cm diameter vacuum chamber is equipped
with a high-power closed-cycle cryocooler
(Advanced Research Systems),
which cools down the sample holder to
a temperature of $5$~K in normal operation mode.
To provide homogeneous ice deposition
(over the optical aperture of a substrate)
for the experiments,
in this work, the radiation shield
was removed from the sample holder, reaching
a minimum achievable temperature of $11$~K. 
A base pressure of $\simeq 10^{-7}$~mbar
is set when the system is cold
via a pumping station composed of
a turbomolecular pump
combined with a backing rotary pump.

The substrate is placed in the middle of the vacuum chamber
and the optical windows of the chamber
are made of
the high-resistivity float-zone silicon (HRFZ-Si)
with the high refractive index of
$n_\mathrm{Si} \approx 3.4$,
negligible dispersion,
and small absorption
in the desired THz--IR range.
It is worth noting that the
broadband dielectric response of
this THz--IR optical material
is almost independent of temperature.

The
TPS spectrometer (BATOP TDS-$1008$)
features a broad spectral range
of $0.05$--$3.5$~THz, with a maximum at $\approx 1.0$~THz
and a spectral resolution as high as $\approx 0.03$~THz.
In this TPS system, a pair of photoconductive antennas
are pumped and probed by
a femtosecond fiber laser (TOPTICA),
serving as an emitter and a detector of THz pulses.
A customized sample compartment
allocates the cryocooler, while the TPS housing is kept
under cold nitrogen gas purging, aimed at suppressing
an impact of atmospheric water along the beam path
on the measured THz data.

A Bruker IFS $125$HR FTIR spectrometer is used to record
the transmission IR spectra of ice,
with a resolution as high as $\simeq 1$~cm$^{-1}$
($\simeq 0.03$~THz).
A Mylar Multilayer beam splitter,
an FIR-Hg source and an FIR-DTGS detector
were selected to work in the FIR and mid-IR ranges.
A customized flange accommodates the cryocooler
in the spectrometer sample compartment,
which is kept under vacuum.

\subsection{The experimental protocol}
\label{SEC:IcePrep}

\begin{figure*}[!t]
    \centering
    \includegraphics[width=2.0\columnwidth]{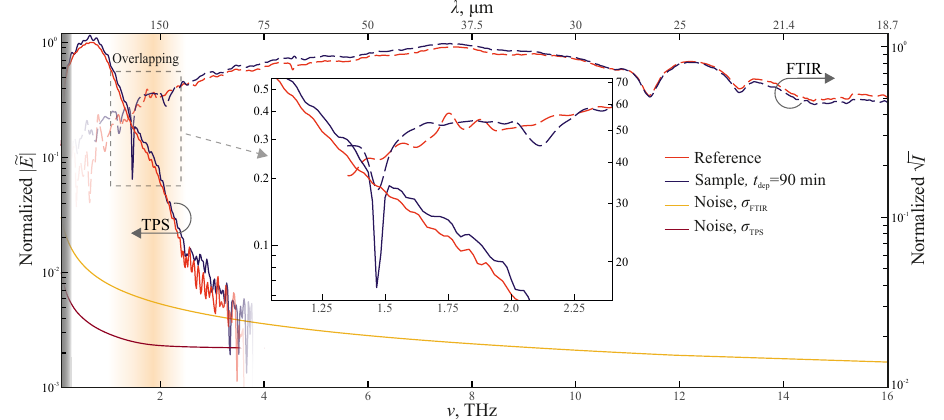}
    \caption{Reference and sample spectra of \ce{N2} ice,
    measured by the TPS (solid lines)
    and FTIR (dashed lines) spectrometers
    at specified deposition steps, $t_\mathrm{dep}$,
    and normalized by the maximum of
    the corresponding reference spectrum
    (for convenience).
    The low-frequency gray-shaded area
    shows the spectral range
    where distortions are expected
    owing to the THz beam diffraction
    at the sample aperture
    \citep{AA.629.A112.2019}.
    The orange-shaded area near $\simeq 2.0$~THz 
    (enlarged in the inset for clarity)
    indicates where the TPS and FTIR data overlaps.
    Sensitivity of the TPS    
    and FTIR measurements is characterized by
    the standard deviation of the corresponding
    instrumental noise, $\sigma_\mathrm{TPS}$ and $\sigma_\mathrm{FTIR}$
    as described by \cite{2022A&A...667A..49G}.}
    \label{FIG:SampleRefSp}
\end{figure*}

The ice samples for the TPS and FTIR measurements were grown under the same experimental conditions. Following the experimental procedure by \citet{AA.629.A112.2019}, the ice layers are formed by the
condensation from the gas phase through a $6$-mm-diameter stainless steel pipe placed at $\simeq 7$~cm from the substrate to ensure uniform deposition on both sides.
A metering valve regulated the \ce{N2} flux. To minimize \ce{H2O} and \ce{CO2} contamination in the gas flow entering the vacuum chamber, the last part of the gas line includes a trap, made of a copper coil and 
immersed in liquid nitrogen.
Such a trap reduces the total gas flux
and, thus, increases by $\simeq 3$ times
the required deposition time
to reach the same ice thickness
as reported in our earlier studies
\citep{AA.629.A112.2019, 2022A&A...667A..49G, AA.scattering.2025}.

A thickness on the order of a fraction of a millimeter
is required to obtain a reliable reconstruction of
the THz--IR optical constants,
so the chosen deposition rate
must be fast enough to grow thick ice
in a reasonable experimental time.
To achieve this condition, the pressure was maintained at $\approx 10^{-4}$~mbar, while each deposition step was set to 9~min, minimizing heating of the HRFZ-Si substrate due to gas condensation. The total deposition time was 90~min, resulting in an overall ice thickness of approximately 1.72~mm. The ice layers on the front and back sides of the substrate reached thicknesses of about 1.028~mm and 0.689~mm, respectively. The substrate temperature was maintained at 11~K prior to deposition and increased slightly to 12~K during each deposition cycle. Between consecutive steps, the system was allowed to equilibrate until the substrate temperature returned to $11$~K.

\subsection{TPS and FTIR data processing}

For both the TPS and FTIR measurements, reference
spectra of the bare substrate
were recorded before the start of the deposition process.
The sample signals were measured
every second deposition step after 
$18$~min. The Tukey \citep{Tukey.1986} and $4^\mathrm{th}$-order Blackman-Harris \citep{ProcIEEE.66.1.51.1978}
apodization windows were applied for pre-processing of
the TPS waveforms and FTIR interferograms, respectively. 
For the TPS data, the application of the Fourier transform
results in the frequency-domain complex amplitude
$\widetilde{E} \left( \nu \right)$,
with both the amplitude and phase information,
which makes possible the direct reconstruction of
the THz complex dielectric permittivity of an analyte
\citep{AA.629.A112.2019}.
In turn, FTIR gives the power spectrum
$I \left( \nu \right) \propto | \widetilde{E} \left( \nu \right) |^{2}$
with only the amplitude information
\citep{GriffithsFTIR_Book1986}.

From the reference (bare substrate) and sample measurements we define the experimental transmission coefficients of the ice films. 
For TPS, where the complex electric-field spectrum $\widetilde{E}(\nu)$ is available, the complex transmission amplitude reads

\begin{equation}
\widetilde{T}_{\mathrm{exp,TPS}}(\nu)
=\frac{\widetilde{E}_{\mathrm{sample}}(\nu)}{\widetilde{E}_{\mathrm{reference}}(\nu)}\,,
\label{eq:T_TPS}
\end{equation}

whereas for FTIR, which provides intensities, we report the amplitude of the transmission as

\begin{equation}
\bigl|\widetilde{T}_{\mathrm{exp,FTIR}}(\nu)\bigr|
=\sqrt{\frac{I_{\mathrm{sample}}(\nu)}{I_{\mathrm{reference}}(\nu)}}\,.
\label{eq:T_FTIR}
\end{equation}

These definitions link the complex transmission of Fig.~2 to the $E$– and $I$–spectra in Fig.~1. In the next step, the TPS and FTIR datasets are merged by (i) reconstructing the FTIR phase via the Kramers–Kronig transform and (ii) matching amplitudes and phases to the TPS reference, following the procedure detailed in \citet{2022A&A...667A..49G}.

\begin{figure}[!t]
    \centering
    \includegraphics[width=1.0\columnwidth]{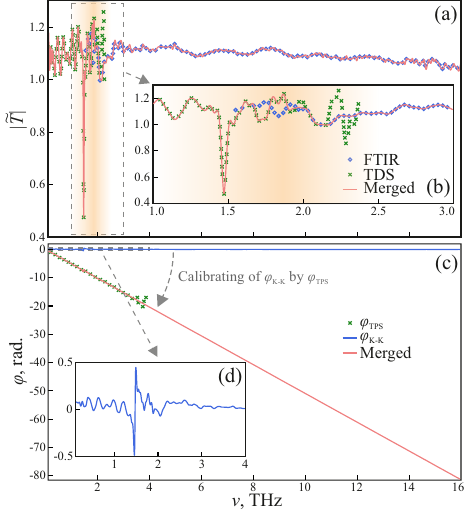}
    \caption{Merging of the TPS and FTIR data
    for N$_2$ ice after the $54$-min-long deposition.
    (a)~Amplitude of the complex transmission coefficient
    $\left| \widetilde{T} \left( \nu \right) \right|$
    retrieved from the TPS (green markers)
    and FTIR (blue) data, and the resultant merged curve (red) see Eqs.~(\ref{eq:T_TPS})--(\ref{eq:T_FTIR}) for the definitions of the transmission coefficients.
    (b)~Zoom-in on the TPS and FTIR overlapping data.
    (c)~Phase of the complex transmission coefficient
    $\phi\left( \nu \right)$, where the
    TPS phase $\phi_\mathrm{TPS}$ (green markers),
    FTIR-based Kramers-Kronig phase $\phi_\mathrm{K-K}$
    (blue line), and resultant broadband phase (red) are shown.
    (d)~Low-frequency behavior of the FTIR-based Kramers-Kronig phase
    $\phi_\mathrm{K-K}$.}
    \label{FIG:MergedTransmission}
\end{figure}

In Fig.~\ref{FIG:SampleRefSp},
the reference and sample spectra
are acquired by TPS and  FTIR systems
for the \ce{N2}-ice samples of
different thicknesses.
The shaded spectral range
of the TPS and FTIR data overlapping
near $\simeq 2.0$~THz
is broad enough to make merging the spectral data possible.
We have no absorption lines except at 1.5 and 2.13 THz.
For both the TPS and FTIR systems,
in Fig.~\ref{FIG:SampleRefSp},
the spectral noise levels are estimated
in the form of frequency-dependent standard deviations
$\sigma_\mathrm{TPS} \left( \nu \right)$
and $\sigma_\mathrm{FTIR} \left( \nu \right)$,
to highlight the variability of
the reference spectra as detailed by
\cite{2022A&A...667A..49G}.
Fig.~\ref{FIG:MergedTransmission}~(a),(b) illustrates the merging of the TPS and FTIR data for the \ce{N2} ice after a $54$-min-long deposition.
In the overlapping range, the measured TPS (green markers)
and FTIR (blue markers) data are close. 
The weighted superposition of
the transmission spectra
in the overlapping range
is used to calculate
the broadband transmission amplitude
$| \widetilde{T} \left( \nu \right) |$,
while the spectral signal-to-noise ratios of our spectrometers
ensure a smooth transition
from the lower to higher frequencies
as shown by \cite{2022A&A...667A..49G}.
In Fig.~\ref{FIG:MergedTransmission}~(c),
the missing IR phase
is retrieved from the FTIR data
using the Kramers-Kronig transform
\citep{PR.161.1.143.1967, AIPAdv.2.3.032144.2012},
while a knowledge of the low-frequency TPS
phase allows for mitigating
the Kramers-Kronig transform uncertainty
\citep{2022A&A...667A..49G}.

In this way broadband complex transmission spectra 
$\widetilde{T} \left( \nu \right)$
was obtained from the experiments
and used to retrieve
both the thicknesses of \ce{N2} ice samples 
and their complex refractive index
$\widetilde{n} = n - i \alpha c_\mathrm{0} / \left( 2 \pi \nu \right)
\equiv \sqrt{ \widetilde{ \varepsilon } }$
in the frequency range of $0.3$--$16.0$~THz, where
$n$ is a real refractive index,
$\alpha$ is an absorption coefficient (by field)\footnote{We note that the power absorption coefficient
(within the radiative transfer theory)
is given by $\mu_\mathrm{\alpha} = 2 \alpha$. },
$c_\mathrm{0} = 3 \times 10^{8}$~m/sec
is the speed of light in vacuum, while
$\widetilde{\varepsilon} = \varepsilon' - i \varepsilon''$
is the complex dielectric permittivity.
For this, we use the earlier-reported approach
\citep{AA.629.A112.2019, 2022A&A...667A..49G}
that minimizes the error functional
-- i.e., a discrepancy between
the experimental complex transmission spectrum
$\widetilde{T} \left( \nu \right)$
and its theoretical model,
which accounts for all key features of
the radiation--sample interactions:
reflections at interfaces,
absorption and phase delay in bulk,
standing waves in layers,
finite coherence length,
and signal apodization.
Moreover, determining the position of
satellite pulses from the TPS waveforms is used to
find the thicknesses of the ice layers
\citep{AA.629.A112.2019}.

\subsection{Theoretical analysis}
\label{SEC:DielModel}

\subsubsection{Analytical model of complex dielectric permittivity}

Resonant absorption peaks observed in
the broadband dielectric spectra of
the \ce{N2} ice were modeled by
a superposition of the Lorentz kernels
\begin{equation}
    \widetilde{\varepsilon}\left( \nu \right)
    = \varepsilon_\infty
    + \sum_{j=1}^{2} \frac{ \Delta\varepsilon_j\nu_{\mathrm{L},j}^{2} }
    { \nu^{2}_{\mathrm{L},j} - \nu^{2}  + i \nu \gamma_{\mathrm{L},j} }\:,
    \label{EQ:LorentzModel}
\end{equation}
where $\Delta\varepsilon_j$,
$\nu_{\mathrm{L},j}$,
and $\gamma_{\mathrm{L},j}$
are the amplitude,
resonant frequency,
and damping constant
of the $j^\mathrm{th}$ Lorentz term,
while $\varepsilon_\infty$
is the constant dielectric permittivity
at higher frequencies.
Parameters $\nu_{\mathrm{L},j}$
and $\gamma_{\mathrm{L},j}$
indicate the spectral position and width of
each Lorentz oscillator,
while $\Delta \varepsilon_j$
is responsible for its contribution 
to the total dielectric response.
It is worth noting that this model
is physically-rigorous since it
satisfies the sum rule
\citep{PR.161.1.143.1967,OE.30.6.9208.2022}
and the Kramers-Kronig relations
\citep{PR.161.1.143.1967, AIPAdv.2.3.032144.2012}.
This favorably distinguishes
the model defined by Eq.~\eqref{EQ:LorentzModel}
from a series of Gaussian bands
often used to analyse
the astrophysical ice absorption properties
\citep{ARAA.52.1.541.2015}.

The first approximations for
the model parameters are estimated 
from the reconstructed dielectric response,
including positions and width of
the absorption peak,
where the magnitude of the peak
can be estimated from
the real dielectric permittivity.
Owing to the convenience of
presenting results via $n(\nu)$ and $\alpha(\nu)$,
this form was chosen for
the measured data representation; if needed, this can be easily converted to $\varepsilon'$ and $\varepsilon''$,
as detailed above.

\subsubsection{DFT calculations of vibrational modes}
\label{SEC:DFT}

To simulate the far-IR spectra, the $\alpha$-phase of the N$_2$ crystal was considered, as expected in our experimental conditions \citep{Wyckoff1963}. It features the primitive cubic crystalline lattice, characterized by four molecules per unit cell of the volume of 179.79~\AA$^3$ and the period of 5.64~\AA.
Such a unit cell was repeated $2\times2\times2$ times, and the gas-phase cluster model was created by removing the N atoms at the boundaries. This allows us to model solid materials with high accuracy by using gas-phase codes, enabling the use of a higher-level theory.

To compute the IR spectra in the THz region (see Appendix~\ref{Theoretical Benchmark}), we employed DFT implemented in the ORCA code \citep{Neese_2020}. We used the B3LYP exchange-correlation functional \citep{B3LYP_cit1,B3LYP_cit2} and a triple-zeta valence basis set with polarization functions Def2-TVZP \citep{Def2-TZVP}.
This method has been reported to reproduce experimental spectra in the THz region of solid-state species belonging to the 2p series of the periodic table \citep{Ugliengo:2021,Kambara,Quiang_2012, Chen_2004,Keith_2010,Dash_2015}. Dispersion correction was not taken into account (see Appendix~\ref{Theoretical Benchmark}), as it causes a redshift of the vibrational modes.

Vibrational frequencies were computed using the harmonic approximation. We opted not to optimize the experimental structure \citep{Wyckoff1963}, because the optimization leads to the larger interatomic distances and, thus, to significant changes in the ab- sorption peak positions. The self-consistent field convergence was achieved when the relative energy change is lower than $10^{-14}$.
The resolution of identity (RI-JONX method) \citep{https://doi.org/10.1002/jcc.10318} approximation was applied to reduce the computational cost for the Coulomb integral only, but not for the Hartree-Fock exchange, in order to ensure the highest possible accuracy at a lower computational cost, using an auto-generated auxiliary basis set \citep{https://doi:10.1021/acs.jctc.6b01041} by ORCA.

\section{Results}
\label{SEC:Results}

\begin{figure}[!b]
    \centering
    \includegraphics[width=1.0\columnwidth]{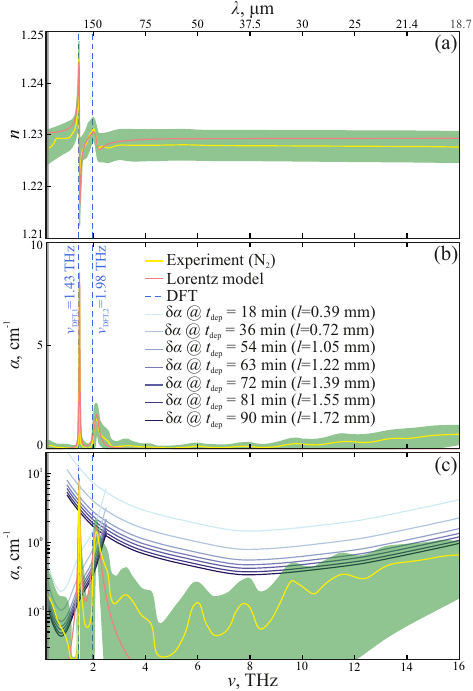}
    \caption{Broadband THz--IR optical properties
    of the \ce{N2} ice,
    deduced from the measurements of sample
    of different thicknesses $l$.
    The yellow solid lines show the mean values,
    the green shaded zones, the $\pm 1.5\sigma$ ($87$\%)
    confidence intervals of the measurements,
    while the red solid lines, the complex dielectric permittivity model
    defined by Eq.~\ref{EQ:LorentzModel} and Table~\ref{TAB}.
    (a)~Refractive index $n$.
    (b,~c)~Absorption coefficient $\alpha$
    (by field)
    in the linear and logarithmic scales,
    respectively.
    In~(c), the blue-shaded solid lines
    define the $3\sigma$ detection limits
    for the absorption $\delta \alpha$, 
    calculated for
    the different ice thicknesses $l$
    (measured with
    the $\sim0.01$~mm uncertainty).
    In~(a)--(c), the vertical blue dashed lines define
    the frequencies of
    the $\alpha$-\ce{N2} vibrational modes
    predicted by the DFT method.}
    \label{FIG:IceResponse}
\end{figure}

In Fig.~\ref{FIG:IceResponse}, the estimated broadband dielectric response of \ce{N2} ice deposited at 
$11$~K is shown as the refractive index $n$
and absorption coefficient $\alpha$ by field).
Yellow curve and green-shaded areas
stand for the average values and
$\pm 1.5 \sigma$ (or $87\%$) confidence intervals,
calculated from the ensemble of measurements
at the different deposition times
in the $18$--$90$~min range
with the $18$~min step.
As shown by \cite{2022A&A...667A..49G}, the
sensitivity of our spectrometers significantly impacts
the reconstruction accuracy,
especially in the case of a low-absorbing species.
Therefore, in Fig.~\ref{FIG:IceResponse}~(c), approximations for
the detection limit are calculated from
the TPS and FTIR noises
($\sigma_\mathrm{TPS}$ and $\sigma_\mathrm{FTIR}$,
shown in Fig.~\ref{FIG:SampleRefSp})
as a function of the deposition time $t_\mathrm{dep}$
(or the total sample thickness $l$)
as detailed in \citep{2022A&A...667A..49G}.
From these estimates,
one notices that two absorption peaks at lower frequencies
are above the detection limit,
while all the IR background absorption is below.
Thus, we further analyze only these two
physically-reasonable absorption peaks
while neglecting the background that is more likely attributed to noise.

In Fig.~\ref{FIG:IceResponse},
the complex dielectric permittivity model
defined by Eq.~\eqref{EQ:LorentzModel}
is estimated and overlapped with
the experimental curves,
while the resultant model parameters
are summarized in Table~\ref{TAB}.
The absorption peaks observed at $1.47$ and $2.13$~THz
agree with the literature data
on studies of the solid-state nitrogen
-- i.e., the molecular crystal,
the THz--IR response of which depends
on the crystalline structure
\citep{JCP.46.10.3991.1967, JCP.45.11.4359.1966, Louis1969, Scott1976, Savchenko2019}.

As discussed in Sec.~\ref{SEC:DFT}, the applied growth conditions 
presume that the \ce{N2} ice is in the $\alpha$ phase.
In such a case, the factor group analysis predicts
$24$ normal modes for the Pa3 space group,
among which only three are translational phonon modes $T_\mathrm{u}$,
including the single acoustic and two
optically active $T_\mathrm{u}$ phonons.
These modes were analyzed theoretically
and studied experimentally
via the FTIR and neutron scattering techniques
by several groups; for example, see
\citep{JCP.46.10.3991.1967, JCP.45.11.4359.1966, Louis1969, Kjems1975}.
The two optical features
and the associated absorption bands show some variations between different experiments. 
For instance, by comparing our results with those of
\cite{JCP.46.10.3991.1967}, we obtain a $\approx 30\%$ discrepancy for
the ratio of the amplitude absorption values  
($0.35$ in \citet{JCP.46.10.3991.1967} versus $0.27$ in the present study)
and the ratio of the integrated absorption intensities 
($0.76$ in \citet{JCP.46.10.3991.1967} versus $1.11$ in our work).

Broadening and some asymmetry of the $2.13$~THz peak
(as compared to the Lorentz model)
might indicate anharmonicity effects,
which can be explained by the dependence between
the mode frequency and damping \citep{Hardy1982},
or the morphological features of ice
(porosity, disorder, and polycrystallinity).
A somewhat monotonic increase in the \ce{N2}-ice absorption
at higher frequencies ($> 8.0$~THz)
is attributed to the scattering on heterogeneities of
an analyte (such as pores,
as reported by \cite{AA.scattering.2025}).
This effect indirectly evidences the imperfect ice structure
but is still below the defined detection limit.

Our DFT calculations of the $\alpha$-N$_2$ vibrational modes depict two harmonic frequencies, representing the expected translational movements of N$_2$ molecules and pointed out in Fig.~\ref{FIG:IceResponse}. From Table~\ref{tab:Freq}, we also notice that these vibrational modes agree with both our and earlier-reported experiments, as well as the theoretical predictions based on the interatomic potentials \citep{Zunger_1975}. The discrepancies are quite small -- at the level of a few inverse centimeters. They are in close agreement within the DFT accuracy \citep{Banks2020}. 
Our theoretical analysis not only reproduced the $\alpha$-N$_2$ translational modes, but also revealed some temperature dependence of the lattice structure and the resultant evolution of vibrational modes with temperature (see Appendix~\ref{Theoretical Benchmark}). However, this aspect is beyond the scope of this article. Thereby, both the literature survey on the response of solid nitrogen and the DFT simulations justify the origin of the N$_2$-ice vibrational modes observed in our THz–IR spectroscopic experiments.

\begin{table}[!t]
    \caption{Parameters of the complex
    dielectric permittivity model of
    \ce{N2} ice defined by Eq.~\eqref{EQ:LorentzModel},
    with $\pm 1\sigma$ confidence intervals.}
    \centering 
        \begin{tabular}{ >{\centering\arraybackslash}m{3.5cm}
                        |>{\centering\arraybackslash}m{3.5cm}}
            \rule[0ex]{0pt}{1.0ex}    {Parameter}
                                     & {Value}\\
            \hline \hline
            {$\varepsilon_\infty$}
                                     & {$1.51 \pm 0.01$}\\
            {$\Delta\varepsilon_\mathrm{1}$}
                                     & {$(16 \pm 1) \times 10^{-4}$}\\
            {$\nu_\mathrm{L,1}$, THz}
                                     & {$1.47 \pm 0.01$}\\
            {$\gamma_\mathrm{L,1}$, THz}
                                     & {$0.03 \pm 0.01$}\\
            {$\Delta\varepsilon_\mathrm{2}$}
                                     & {$(9.6\pm 2.2) \times 10^{-4}$}\\
            {$\nu_\mathrm{L,2}$, THz}
                                     & {$2.13 \pm 0.02$}\\
            {$\gamma_\mathrm{L,2}$, THz}
                                     & {$0.22 \pm 0.03$}\\
           \end{tabular}
    \label{TAB}
\end{table}

\begin{table*}[!t]
    \caption{Optically-active vibrational modes of
    the $\alpha$-\ce{N2} crystal
    found experimentally and calculated via DFT,
    as compared to some previous experimental and theoretical studies.
    Here, $k \equiv 1/ \lambda = \nu / c_\mathrm{0}$
    stands for the spectroscopic wavenumber.}
    \centering 
        \begin{tabular}{ >{\centering\arraybackslash}m{6.0cm}
                        |>{\centering\arraybackslash}m{5.0cm}
                        |>{\centering\arraybackslash}m{5.0cm}}
            \rule[0ex]{0pt}{1.0ex}    {Method}
                                     & {$\nu_1$, THz
                                     ($k_1$, cm$^{-1}$)}
                                     & {$\nu_2$, THz ($k_2$, cm$^{-1}$)}\\
            \hline \hline
            {Exp. (this work)}
                                     & {$1.47$ ($49$)}
                                     & {$2.13$ ($71$)}\\
            {DFT (this work)}
                                     & {$1.43$ ($47.57$)}
                                     & {$1.98$ ($66.07$)}\\        
            {Exp. @ $10$~K \citep{Anderson_1966}}
                                     & {$1.44 \pm 0.12$ ($48 \pm 4$)}
                                     & {$2.07 \pm 0.21$ ($69 \pm 7$)}\\
            {Exp. @ $15$~K \citep{Kjems1975}}
                                     & {$1.45 \pm 0.02$ ($48.4 \pm 0.8$)}
                                     & {$2.08 \pm 0.05$ ($69.4 \pm 1.6$)}\\
            {Exp. @ $20$~K \citep{JCP.46.10.3991.1967}}
                                     & {$1.47 \pm 0.09$ ($49 \pm 3$)}
                                     & {$2.07 \pm 0.18$  ($69 \pm 6$)}\\
            {Int. potentials \citep{Zunger_1975}}
                                     & {$1.41$ ($47$)}
                                     & {$2.08$ ($69.5$)}
        \end{tabular}
    \label{tab:Freq}
\end{table*}

\section{Discussion and conclusions}

In this work, we combined TPS and FTIR spectroscopy
to study the THz--IR response of
laboratory analogs of \ce{N2} astrophysical ice.
The main results of our findings can be summarized as:
\begin{enumerate}

\item The optical properties of \ce{N2} ice
deposited on the HRFZ-Si substrate at the temperature of $11$~K
were retrieved in the $0.3$--$16$~THz frequency range
or the $1$~mm--$18.75$~$\mu$m wavelength range.
Two resonant absorption peaks of \ce{N2} ice,
those above the detectable absorption level of our spectrometers,
were observed at the frequencies of $1.47$ and $2.13$~THz 
($204$ and $141$~$\mu$m)
and attributed to the optically active translational phonon modes
of the $\alpha$-\ce{N2} molecular crystal.
The $2.1$~THz peak exhibits broadening and asymmetry,
possibly due to anharmonicity effects or morphological features,
such as porosity and polycrystallinity.
A monotonic increase in absorption at higher frequencies ($> 8$~THz), 
although below the detection limit,
is attributed to the scattering on the ice heterogeneities
and, thus, also highlights the imperfections in its structure.

\item The complex dielectric permittivity of
\ce{N2} ice was modeled by a pair of the Lorentz kernels,
which reproduce the observed vibrational modes
and have the resonant frequencies of
$\nu_\mathrm{L} = 1.47$ and $2.13$~THz
and the damping constants of
$\gamma_\mathrm{L} = 0.03$ and $0.22$~THz,
respectively.
The observed resonances
were compared with
the DFT prediction and the available literature data
on the phonon spectrum of $\alpha$-\ce{N2} molecular crystal.
This made it possible to justify
the correctness of
our experimental findings and theoretical interpretations.

\item The density functional theory has predicted 
the translational modes of $\alpha$-N$_{2}$ within 5~cm$^{-1}$ compared to our findings. We have found that translational modes are sensitive to the geometrical structure of $\alpha$-N$_{2}$. Small variations in the interatomic distances, which are temperature-dependent, can drastically change the far-infrared spectra.

\item Our optical constants for \ce{N2} ice cover the far-infrared and submillimeter range, which is poorly sampled by current space-based observatories. While JWST provides high-sensitivity coverage in the near- to mid-infrared, and ALMA operates at longer (submillimeter) wavelengths, a spectral gap remains in the far-infrared (30–200~$\mu$m). Our data help fill this gap from the modeling side. Future instruments such as PRIMA (Probe far-InfraRed Mission for Astrophysics), currently in development, are expected to provide observational access to this range with high spectral resolution and sensitivity. In this context, the dielectric response of solid \ce{N2} obtained in this study provides valuable input for radiative transfer models of N$_2$-rich environments, such as the outer regions of protoplanetary disks.

\end{enumerate}

Finally, accurate optical constants across a broad spectral range are essential for modeling dust and ice in protoplanetary disks. Misestimates of the refractive index can lead to errors in radiative transfer calculations, affecting the inferred disk temperature, density, and composition. Expanding the availability of laboratory measurements helps improve the reliability of disk models and provides deeper insights into the early stages of star and planet formation.

\begin{acknowledgements}
The authors appreciate the support of the Max-Planck Society. F.K. acknowledges support from the European Union’s Horizon Europe research and innovation programme under the Marie Skłodowska-Curie Actions Postdoctoral Fellowship grant agreement No. \#~$101153804$ (ORCHID). This project has also received funding from the European Union’s Horizon~2020 research and innovation program under the Marie Skłodowska-Curie grant agreement \#~$811312$ for the Project "Astro-Chemical Origins" (ACO). TPS and FTIR data processing and analysis
by A.A.G., G.A.K.,  S.V.G., and K.I.Z.
was supported by the RSF~Project
\#~$25$--$79$--$30006$. D.C. acknowledges the Alexander von Humboldt Foundation for funding, as well as support from the University of Perugia with a postdoctoral research fundings (assegno di ricerca), the AIAS–AUFF Fellowship Programme, funded by Aarhus University Research Foundation and Aarhus Institute of Advanced Studies (AIAS) at Aarhus University, and from the Danish National Research Foundation through the Center of Excellence "InterCat" (Grant agreement no.: DNRF150). The authors acknowledges the state of Baden-Württemberg through bwHPC and the German Research Foundation (DFG) through grant no INST 40/575-1 FUGG (JUSTUS 2 cluster).
\end{acknowledgements}

\bibliographystyle{aa} 
\bibliography{REFs}

\appendix

\section{DFT calculations}
\label{Theoretical Benchmark}

\begin{table*}[!t]
\small
    \centering
    \caption{DFT methods (Method)
    before optimization (unrelaxed)
    and after optimization (relaxed),
    vibrational frequencies
    (in the form of
    spectroscopic wavenumbers
    $k \equiv 1/ \lambda = \nu / c_\mathrm{0}$),
    intramolecular distances within
    a nitrogen molecule ($d$(\ce{N-N})),
    and intermolecular distances
    between two nitrogen molecules ($d$(\ce{N2-N2}))
    to simulate an $\alpha$-\ce{N2} cluster model.
    The computed frequencies for librational (L) and translational (not specified) modes are presented. Experimental values (Exp.) and theoretical results based on interatomic potentials (Int. potentials) are also reported.}
        \begin{tabular}{ >{\centering\arraybackslash}m{4.6cm}
                        |>{\centering\arraybackslash}m{9.8cm}
                        |>{\centering\arraybackslash}m{1.2cm}
                        |>{\centering\arraybackslash}m{1.2cm}}
          \rule[0ex]{0pt}{1.0ex}
          Method & $k$, cm$^{-1}$ & $d$(\ce{N2-N2}), \AA & $d$(\ce{N-N}), \AA \\
          \hline\hline
          B3LYP/Def2-SVP (unrelaxed) & $44.68$, $63.97$ & $3.64$ & $1.05$ \\
          B3LYP/pcseg-1 (unrelaxed) &  $38.55$, $56.38$ & $3.64$ & $1.05$ \\
          B3LYP/ano-pvdz (unrelaxed) & $45.45$, $64.06$ & $3.64$ & $1.05$ \\
          B3LYP/Def2-TZVP (unrelaxed) & $47.57$, $66.07$ & $3.64$ & $1.05$ \\
          \hline
          B3LYP-D3/Def2-SVP (unrelaxed) & $40.25$, $55.83$ & $3.64$ & $1.05$ \\
          B3LYP-D3/pcseg-1 (unrelaxed) & $33.34$, $48.13$ & $3.64$ & $1.05$ \\
          B3LYP-D3/ano-pvdz (unrelaxed) & $40.83$, $55.77$ & $3.64$ & $1.05$ \\
          B3LYP-D3/Def2-TZVP (unrelaxed) & $43.18$, $58.14$ & $3.64$ & $1.05$ \\      
          \hline
          B3LYP/pcseg-1 (relaxed) & $21$(L), $28.89$(L), $32.33$(L), $35.58$(L), $45.27$(L) & $3.81$ & $1.1$ \\
          B3LYP/Def2-TZVP (relaxed) & $18.96$, $30.49$(L), $40.98$(L) & $4.07$ & $1.09$ \\
          \hline          
          B3LYP-D3/pcseg-1 (relaxed) & $35.36$(L), $45.63$(L), $53.33$(L), $53.84$(L), $57.16$(L), $60.42$(L), $64.07$(L) & $3.46$ & $1.1$ \\
          B3LYP-D3/Def2-TZVP (relaxed) &  $24.3$(L) , $32.57$(L), $49.95$(L), $54.04$(L), $64.22$(L), $68.75$(L) & $3.61$ & $1.09$ \\
          \hline
          B3LYP/Def2-SVP (relaxed) \tnote{*} & $27.41$(L), $39.23$(L), $39.72$(L), $45.79$(L), $47.09$(L), $54.24$(L) & $3.75$ & $1.1$ \\
          B3LYP-D3/Def2-SVP (relaxed) \tnote{*} & $28.24$(L), $36.71$(L), $50$(L), $60.49$(L), $65.50$(L), $68.95$(L), $90.52$(L), $106.84$(L) & $4.04$ & $1.1$ \\
         \hline
          Exp. (this work)  & $49.0$, $71$ (@ $13$~K) & & \\
          Exp. \citep{Anderson_1966} & $48 \pm 4$, $69 \pm 7$ (@ $10$~K) & & \\
          Exp. \citep{Kjems1975} & $48.4 \pm 0.8$, $69.4 \pm 1.6$ (@ $15$~K) & &  \\
          Exp. \citep{JCP.46.10.3991.1967} & $49 \pm 3$, $69 \pm 6$ (@ $20$~K) & & \\
          Int. potentials \citep{Barbara_1960} & & $4$ & $1.1$ \\
          Exp.\citep{Kjems1975} &  & $1.05$ \\
          Exp. \citep{Kjems1975} &  & $1.01$ \\ [3ex]
    \multicolumn{4}{l}{\small *3x3x3 model.}
   \end{tabular}
    \label{tab:bench_freq}
\end{table*}

Table~\ref{tab:bench_freq} presents
the far-infrared vibrational frequencies of
the $\alpha$-N$_2$ ice cluster model
calculated using the B3LYP functional
with different basis sets
(Def2-SVP \citep{Def2-TZVP},
pcseg-1 \citep{pcseg-1},
ano-pvdz \citep{ano-pvdz},
and Def2-TZVP).
The frequencies are shown
both without (unrelaxed)
and with (relaxed) optimization,
except for the ano-pvdz basis set,
which encountered convergence issues during optimization.

Without optimizing the structure,
the Def2-TZVP basis set shows smaller deviations
($1.43$ and $4.93$~cm$^{-1}$)
with respect to the experimental values,
while the pcseg-1 basis set shows
the highest deviations
($10.45$ and $14.62$~cm$^{-1}$).
When adding the Grimme D3 dispersion correction
\citep{Grimme_2006},
the frequencies exhibit a red shift of
$\simeq4.66$~cm$^{-1}$
for the lowest vibrational mode
and $\simeq 8.15$~cm$^{-1}$
for the highest one.

When optimizing the experimental structure
to a local minimum,
the N$_2$ intramolecular distances vary
depending on the basis set,
while the intermolecular \ce{N-N} distances
do not deviate significantly.
After optimization,
we have noticed that
the far-infrared spectrum completely changes,
giving rise to infrared-active librations
(oscillations along the center of mass of
a diatomic molecule).
This suggests that the far-infrared region
is strongly sensitive to
the \ce{N2} interatomic distances.
Hence, depending on the temperature,
translational modes might vary or even disappear,
giving rise to librational modes.

\begin{figure}[!h]
    \centering
    \includegraphics[width=\columnwidth]{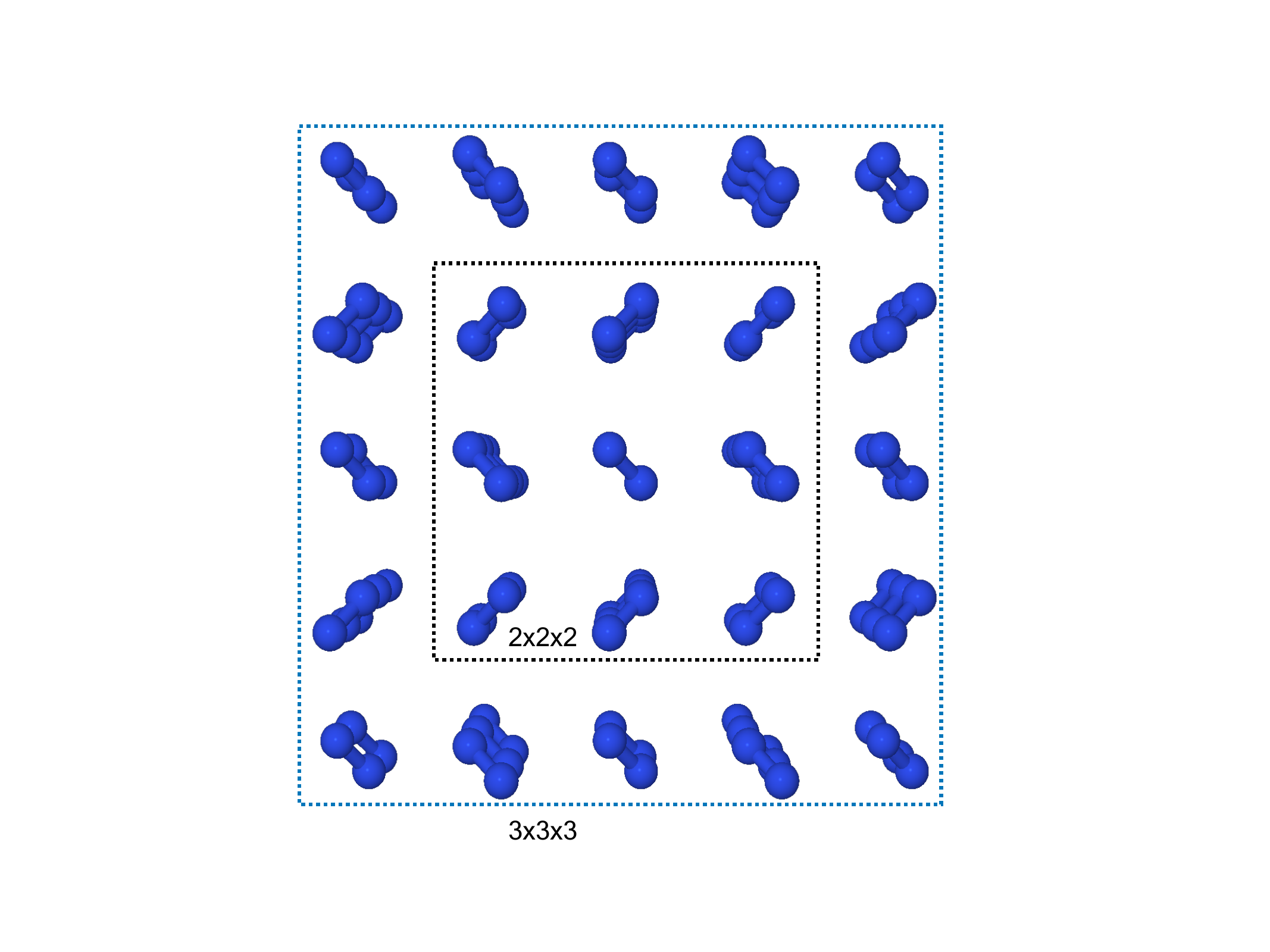}
    \caption{The balls and sticks model of
    the $3 \times 3 \times 3$ cluster of
    $\alpha$-\ce{N2} ice
    is derived by cutting
    a $3 \times 3 \times 3$ repeated bulk
    using the same procedure adopted for obtaining
    the $2 \times 2 \times 2$ cluster model, 
    which is also schematically reported here
    for comparison.}
    \label{Fig:Cluster_model_3x3x3}
\end{figure}

We expanded our cluster model
by cutting it from
a $3 \times 3 \times 3$ bulk of $\alpha$-\ce{N2}
(Fig.~\ref{Fig:Cluster_model_3x3x3})
and optimized it to evaluate the effect of size on
reproducing the vibrational modes.
Due to the large number of atoms,
we were limited to using B3LYP/Def2-SVP
due to computational constraints
and only computed the partial Hessian numerically,
making vibrational calculations only for
the atoms within the $2 \times 2 \times 2$ box.
Optimization using B3LYP-D3
leads to the appearance of librational modes,
as the dispersion correction in this case
underestimates intramolecular bond distances
by $\simeq 0.35$~$\AA$
compared to B3LYP without dispersion.

\end{document}